\documentclass[aps,prb,10pt,amsmath,twocolumn,amssymb,superscriptaddress,showpacs]{revtex4-1} 
\usepackage{bm}
\usepackage[]{graphicx}
\usepackage[tight]{units}
\usepackage{color}
\usepackage{ulem}
\usepackage{marvosym}
\usepackage[colorlinks=true,citecolor=blue,linkcolor= blue,linkcolor=blue,urlcolor=blue]{hyperref}

\definecolor{lred}{rgb}{.55,.0,.0}

\newcommand{\cip}{$\sigma_{_{\|}}$~}
\newcommand{\cpp}{$\sigma_{_{\perp}}$~}

\newcommand{\cppx}{$\sigma_{_{\perp}}$}
\newcommand{\PFip}{$\text{PF}_{_{\|}}$ ~}
\newcommand{\PFpp}{$\text{PF}_{_{\perp}}$~}

\newcommand{\PFipx}{$\text{PF}_{_{\|}}$}

\newcommand{\Sip}{$S_{_{\|}}$~}

\newcommand{\ratiox}{$\nicefrac{\sigma_{\|}}{\sigma_{\perp}}$}

\newcommand{\ratioPFx}{$\nicefrac{\text{PF}_{\|}}{\text{PF}_{\perp}}$}

\newcommand{\ratio}{$\nicefrac{\sigma_{\|}}{\sigma_{\perp}}$~}
\newcommand{\ratioS}{$\nicefrac{S_{\|}}{S_{\perp}}$~}
\newcommand{\ratioPF}{$\nicefrac{\text{PF}_{\|}}{\text{PF}_{\perp}}$~}

\newcommand{\Sppx}{$S_{_{\perp}}$}
\newcommand{\BiTe}{$\text{Bi}_2\text{Te}_3$~}
\newcommand{\SbTe}{$\text{Sb}_2\text{Te}_3$~}
\newcommand{\SBSL}{$\text{Bi}_2\text{Te}_3/\text{Sb}_2\text{Te}_3$ SL~}
\newcommand{\SB}{$\text{Bi}_2\text{Te}_3/\text{Sb}_2\text{Te}_3$~}

\newcommand{\BiTex}{$\text{Bi}_2\text{Te}_3$}
\newcommand{\SbTex}{$\text{Sb}_2\text{Te}_3$}
\newcommand{\SBSLx}{$\text{Bi}_2\text{Te}_3/\text{Sb}_2\text{Te}_3$ SL}

\newcommand{\SBxSL}{$(\text{Bi}_2\text{Te}_3)_{x}/(\text{Sb}_2\text{Te}_3)_{1-x}$ SL~}

\newcommand{\SBSLsx}{$\text{Bi}_2\text{Te}_3/\text{Sb}_2\text{Te}_3$ SLs}
\newcommand{\SBSLs}{$\text{Bi}_2\text{Te}_3/\text{Sb}_2\text{Te}_3$ SLs~}

\newcommand{\SBxSLs}{$(\text{Bi}_2\text{Te}_3)_{x}/(\text{Sb}_2\text{Te}_3)_{1-x}$ SLs~}
\newcommand{\f}[1]{Fig.~\ref{fig:#1}}
\newcommand{\fs}[1]{Figs.~\ref{fig:#1}}
\newcommand{\F}[1]{Figure \ref{fig:#1}}

\newcommand{\Sec}[1]{Sec.~\ref{#1}}

\newcommand{\V}{Venkatasubramanian \textit{et al.}~\cite{Venkatasubramanian:2001p114}~}
\newcommand{\Vx}{Venkatasubramanian \textit{et al.}~\cite{Venkatasubramanian:2001p114}}

\newcommand{\Vno}{Venkatasubramanian \textit{et al.}~}

\newcommand{\Vline}{Ref.~\onlinecite{Venkatasubramanian:2001p114}~}
\newcommand{\Vlinex}{Ref.~\onlinecite{Venkatasubramanian:2001p114}}

\begin{document}

\title[]{Thermoelectric transport in $\text{Bi}_2\text{Te}_3/\text{Sb}_2\text{Te}_3$ superlattices}
\author{N. F. Hinsche}
\email{nicki.hinsche@physik.uni-halle.de}
\affiliation{Institut f\"{u}r Physik, Martin-Luther-Universit\"{a}t Halle-Wittenberg, DE-06099 Halle, Germany}
\author{B. Yu. Yavorsky}
\affiliation{Forschungszentrum Karlsruhe, Herrmann-von-Helmholtz-Platz 1, 76344 Eggenstein-Leopoldshafen, Germany.}
\author{M. Gradhand}
\affiliation{H. H. Wills Physics Laboratory, University of Bristol, Bristol BS8 1TH, United Kingdom}
\author{M. Czerner}
\affiliation{I. Physikalisches Institut, Justus-Liebig-Universit\"{a}t Gie\ss en, DE-35392 Gie\ss en, Germany}
\author{M. Winkler}
\affiliation{Fraunhofer Institut f\"{u}r Physikalische Messtechnik, Heidenhofstrasse 8, DE-79110 Freiburg, Germany}
\author{J. K\"{o}nig}
\affiliation{Fraunhofer Institut f\"{u}r Physikalische Messtechnik, Heidenhofstrasse 8, DE-79110 Freiburg, Germany}
\author{H. B\"{o}ttner}
\affiliation{Fraunhofer Institut f\"{u}r Physikalische Messtechnik, Heidenhofstrasse 8, DE-79110 Freiburg, Germany}
\author{I. Mertig}
\affiliation{Institut f\"{u}r Physik, Martin-Luther-Universit\"{a}t Halle-Wittenberg, DE-06099 Halle, Germany}
\affiliation{Max-Planck-Institut f\"{u}r Mikrostrukturphysik, Weinberg 2, DE-06120 Halle, Germany}
\author{P. Zahn}
\affiliation{Helmholtz-Zentrum Dresden-Rossendorf, P.O.Box 51 01 19, DE-01314 Dresden, Germany}
%
\date{\today}

\begin{abstract}
 The thermoelectric transport properties of $\text{Bi}_2\text{Te}_3/\text{Sb}_2\text{Te}_3$ 
 superlattices are analyzed on the basis of first-principles calculations and semi-classical Boltzmann theory. 
 The anisotropy of the thermoelectric transport under electron and hole-doping was studied in 
 detail for different superlattice periods at changing temperature and charge carrier concentrations.
 A clear preference for thermoelectric transport under hole-doping, as well as for the in-plane transport 
 direction was found for all superlattice periods. At hole-doping the electrical transport anisotropies 
 remain bulk-like for all investigated systems, while under electron-doping quantum confinement 
 leads to strong suppression of the cross-plane thermoelectric transport at several superlattice periods. 
 In addition, insights on the Lorenz function, the electronic contribution to the thermal conductivity and 
 the resulting figure of merit are given.
\end{abstract}

\pacs{31.15.A-,71.15.Mb,72.20.Pa,72.20.-i}

\maketitle


\section{Introduction}
Solid-state thermoelectric (TE) power generation devices 
possess the desirable nature of being highly reliable, stable, compact and integrable 
and have potential applications in waste-heat recovery and outer space explorations. 
However, while intensively studied in the last decades, poor energy 
conversion efficiencies below a few percent at room temperature prohibited the triumph 
of the TE devices as promising alternative energy sources. 
The conversion performance of a thermoelectric material is quantified by the 
figure of merit (FOM) 
\begin{equation}
ZT=\frac{\sigma S^{2}}{\kappa_{el} + \kappa_{ph}} T,
\label{eq1}
\end{equation}
where $\sigma$ is the electrical conductivity, $S$ the thermopower, $\kappa_{el}$  and 
$\kappa_{ph}$ are the electronic and lattice contribution to the thermal conductivity, respectively. 
From Eq.~\ref{eq1} it is obvious, that a higher $ZT$ is obtained by decreasing the denominator 
or by increasing the numerator, the latter being called power factor $\text{PF}=\sigma S^{2}$. 
While $\sigma$, $S$, $\kappa_{el}$ and $\kappa_{ph}$ can individually be tuned by several
orders of magnitude, the interdependence between
these properties impede high values for the FOM \cite{Snyder:2008p240,Vineis:2010p14995}. 
\BiTex, \SbTe and their related 
alloys dominate the field of thermoelectrics with $ZT$ around 
unity from the 1950's through today ~\cite{Wright:1958p15748,Caillat:1992p15526,Poudel:2008p7977}. 

The idea of thermoelectric superlattices (SL) allows for concepts, 
which could enable both, the suppression of 
the cross-plane thermal conductivity \cite{SlackRowe,Chen:1998p15242,Venkatasubramanian:2000p7305} 
and the increase of the electronic power factor \cite{Koga:1999p15774,*Koga:1999p15445,*Koga:1999p5363,*Koga:2000p2542}. 
It suggests that cross-plane transport along the direction 
perpendicular to the artificial interfaces of the SL reduces phonon heat conduction 
while maintaining or even enhancing the electron transport \cite{Bottner:2006p2812}. 
In 2001 a break-trough experiment by \Vno reported a record apparent $ZT = 2.4$ for 
p-type $\text{Bi}_2\text{Te}_3/\text{Sb}_2\text{Te}_3$ and $ZT = 1.4$ for 
n-type $\text{Bi}_2\text{Te}_3/\text{Bi}_2\text{Te}_{2.83}\text{Se}_{0.17}$ superlattices \cite{Venkatasubramanian:1999p13956,Venkatasubramanian:2000p7305,Venkatasubramanian:2001p114}, 
although this values have not yet been reproduced to the best of our knowledge.

With the availability of materials with $ZT \ge 3$ thermoelectric materials could compete with conventional 
energy conversion methods and new applications could emerge \cite{Majumdar:2004p6568}. 
Beside thermal conductivities below the alloy limit, the investigations of \V found a strong decrease 
of the mobility anisotropy and the related electronic thermoelectric 
properties for the SLs at certain periods. 
This is counter-intuitively, as superlattices are anisotropic by definition 
and even the telluride bulk materials show intrinsic anisotropic structural 
and electronic properties \citep{Delves:1961p7491,Stordeur:1976p15137,Stordeur:1975p15016,Huang:2008p559,Hinsche:2011p15707}. 
While considerable effort was done in experimental research
~\cite{Beyer:2002p15267,Bottner:2004p6591,Konig:2011p48,Liao:2010p11008,Peranio:2006p15247,Touzelbaev:2012p15270,Peranio:2011p15471,Winkler:2012p15778}, 
theoretical investigations on \SBSLs are rare. Various available theoretical works concentrate on the electronic structure and 
transport properties of the bulk materials
~\cite{Scheidemantel:2003p14961,Thonhauser:2003p14996,Wang:2007p15244,Huang:2008p559}, 
with some of them discussing the influence of strain, which could occur at the SL interfaces \cite{Wang:2006p15269,Park:2010p11006,Hinsche:2011p15707}. 
To our knowledge, only a sole theoretical work discussed the possible transport 
across such telluride SL structures. Based on density functional theory, Li \textit{et al.}\cite{Li:2004p15238} 
focussed on the calculation of the electronic structure for two distinct \SBSLsx, stating changes 
of the mobility anisotropy estimated from effective masses. 
To extend this work and to clarify the open questions on the reduced mobility anisotropy 
and the enhanced thermoelectric efficiency, 
we are going to discuss in this paper the anisotropic thermoelectric electronic transport 
of seven different \SBSLsx, including the bulk materials,  on the basis of density functional 
theory and semi-classical transport calculations.

\vspace{0.3cm}
For this purpose the paper will be organized as follows. 
In \Sec{method} we introduce our first principle 
electronic structure calculations based on density functional theory and the semi-classical transport calculations 
based on the solution of the linearized Boltzmann equation. 
A brief discussion of the obtained band structures, including the dependence of the band gap 
on different SL periods, is done in section \Sec{estruc}. 
With this, we present in \Sec{unstrained} the electronic thermoelectric 
transport properties, that is electrical conductivity, thermopower and the related power factor, 
of the \SBSLs at different SL periods with a focus on their directional anisotropies. The discussions 
cover a broad temperature and doping range and will conclude the bulk materials. 
Even though only p-type conduction was found in SLs stacked out of pure \BiTe and \SbTe 
n-type conduction will be studied, too, as being possible at appropriate extrinsic doping. 
For the latter case strong quantum confinement effects were found, which will be discussed in detail 
in \Sec{QW}. To give a clue on possible values for the figure of merit, in \Sec{FOM} results for the electronic 
contribution to the thermal conductivity, the Lorenz function as well as existing experimental results 
for the lattice part of the thermal conductivity will be presented. Most of the discussions are done in 
a comparative manner considering the experimental findings of \Vline.

\section{\label{method} Methodology}
\begin{figure}[t]
\centering
\includegraphics[width=0.48\textwidth]{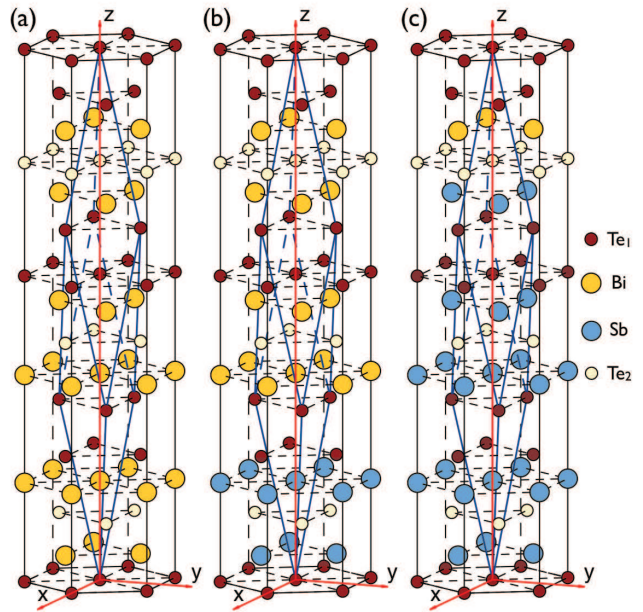}
\caption{\label{fig:1} (color online) Shown are three possible hexagonal unit cells of the 
$(\text{Bi}_2\text{Te}_3)_{x}/(\text{Sb}_2\text{Te}_3)_{1-x}$ superlattices. 
(a) x=1 which is bulk \BiTex, (b) x=${\nicefrac{2}{3}}$ and (c) x=${\nicefrac{1}{6}}$. A concentration 
x=0 would coincide with bulk \SbTe in the lattice of \BiTex.}
\end{figure}
For both bismuth and antimony telluride, as well as for the composed heterostructures, 
we used the experimental lattice parameters and relaxed atomic positions \cite{Landolt} as provided 
for the hexagonal \BiTe crystal structure with 15 atomic layers. 
The layered structure itself is composed out of three formula units, Te$_1$-Bi-Te$_2$-Bi-Te$_1$, 
often called quintuples.  
The hexagonal lattice parameters are chosen to be  
${a^{hex}_{\text{BiTe}}}=4.384${\AA} and $c^{hex}_{\text{BiTe}}=30.487${\AA} for 
\BiTex, \SbTe and the \SBSLsx, respectively. In fact, the main difference between the lattices of 
\BiTe and \SbTe is a decrease of the in-plane lattice constant with an accompanied 
decrease in cell volume. So, a change between the two lattice constants can be 
related to either compressive or tensile in-plane strain. Preceding intense studies revealed 
that a larger in-plane lattice constant, e.g. ${a^{hex}_{\text{BiTe}}} > {a^{hex}_{\text{SbTe}}}$, is favourable for an enhanced cross-plane TE 
transport \cite{Hinsche:2011p15707,Yavorsky:2011p15466,Wang:2007p15244}. For this purpose, the experimental lattice parameters of \BiTe 
were chosen for the studied heterostructures. Structural relaxations revealed only minor influences 
on the bulk electronic structure \cite{Li:2004p15238,Park:2010p11006,Wang:2006p15269,Wang:2007p15244} 
and are beyond the scope of this work. 
To introduce SLs with different layer periods compared to the 
experiments of \V we subsequently substitute the Bi site 
by Sb, starting with six Bi sites in hexagonal bulk \BiTe (see \f{1}(a)). For instance, substituting two atomic layers 
of Bi with Sb leads to a \SBxSL with $x=\nicefrac{2}{3}$, that is two quintuple \BiTe and one quintuple \SbTe (see \f{1}(b)). 
The latter case coincides with a (20\AA/10\AA)-(\BiTex/\SbTex) superlattice 
in the experimental notation of \Vline.

Our thermoelectric transport calculations are performed in two steps. In a first step, the detailed electronic structure of the \SBSLs were 
obtained by first principles density functional theory calculations (DFT), 
as implemented in the fully relativistic screened Korringa-Kohn-Rostoker Greens-function method (KKR) \cite{Gradhand:2009p7460}. 
Within this approach the \textsc{Dirac}-equation is solved self-consistently and with that spin-orbit-coupling (SOC) is included. 
Exchange and correlation effects were accounted for by the local density approximation (LDA) parametrized by Vosco, Wilk, and
Nusair \cite{Vosko1980}. Detailed studies on the electronic structure and transport anisotropy 
of the bulk tellurides \BiTe and \SbTe were published before \cite{Hinsche:2011p15707,Yavorsky:2011p15466} and show very good agreement to 
experimental results and other theoretical findings. 

With the well converged results from the first step we obtain 
the thermoelectric transport properties 
by solving the linearized Boltzmann equation in relaxation time approximation (RTA) within an in-house developed Boltzmann 
transport code \cite{Mertig:1999p12776,Hinsche:2011p15276}. 
Boltzmann transport calculations for thermoelectrics have been carried out for quite a long time and show 
reliable results for wide- and narrow gap 
semiconductors~\cite{Singh:2010p14285,Parker:2010p13171,May:2009p14962,Lee:2011p14982,Hinsche:2011p15276}. 
Calculations on the electronic structure and TE transport 
for bulk \BiTe \cite{Huang:2008p559,Lee:2006p1608,Park:2010p11006,Situmorang:1986p15020} and \SbTe \cite{Thonhauser:2004p15235,Thonhauser:2003p14996,Park:2010p11006} were presented before.
Here the relaxation time $\tau$ is assumed to be isotropic and constant with respect to wave vector k and energy on the scale of $k_{B}T$. 
This assumption is widely accepted for degenerate doped semiconductors. 
Within the RTA, from comparison of the calculated electrical and electronic thermal conductivities (eq.~\ref{Seeb} and \ref{kel}) 
with experiment it is possible to conclude on the relaxation time. 
%
In the following $\tau$ is set to $\unit[10]{fs}$ for 
all bulk and heterostructure systems, regardless of any directional anisotropy or charge 
carrier dependence. We note, that within RTA for the thermopower S (eq.~\ref{Seeb}) 
the dependence of the transport distribution function (TDF), as introduced in the next paragraph, on the energy is essential. 
That is, not only the slope of the TDF, moreover the overall functional behaviour of the TDF on the considered 
energy scale has to change to observe an impact on the thermopower. 

Within the RTA the TDF $\mathcal{L}_{\perp, \|}^{(0)}(\mu, 0)$~\cite{Mahan:1996p508} and with this the 
generalized conductance moments $\mathcal{L}_{\perp, \|}^{(n)}(\mu, T)$ are defined as 
\begin{eqnarray}
& \mathcal{L}_{\perp, \|}^{(n)}(\mu, T)= \nonumber \\
&\frac{\tau}{(2\pi)^3} \sum \limits_{\nu} \int\ d^3k \left( v^{\nu}_{k,(\perp, \|)}\right)^2 (E^{\nu}_k-\mu)^{n}\left( -\frac{\partial f(\mu,T)}{\partial E} \right)_{E=E^{\nu}_k} \nonumber .
\\
\label{Tcoeff}
\end{eqnarray}
$E_k^{\nu}$ denotes the band structure of band $\nu$, $v_k^{\nu}$ the group velocity and $f(\mu,T)$ 
the \textsc{Fermi-Dirac}-distribution. 
$v^{\nu}_{k,(\|)}$, $v^{\nu}_{k,(\perp)}$ denote the group velocities in the directions in the hexagonal basal plane and perpendicular  to it, respectively. 
Within here the group velocities were obtained as derivatives along 
the lines of the Bl\"ochl mesh in the whole Brillouin zone (BZ)~\cite{Yavorsky:2011p15466}. 
The directions of these lines are parallel to the reciprocal space vectors
and so the anisotropy of the real lattice is reflected in these vectors. 
A detailed discussion on implications and difficulties on the numerical determination 
of the group velocities in highly anisotropic materials was currently published elsewhere\cite{Zahn:2011p15523}. 
As can be seen straight forwardly, the temperature- and doping-dependent 
electrical conductivity $\sigma$ and thermopower $S$ 
in the in- and cross-plane directions are defined as
\begin{eqnarray}
\sigma_{_{\perp, \|}}=e^2 \mathcal{L}_{\perp, \|}^{(0)}(\mu, T) \qquad 
S_{_{\perp, \|}}=\frac{1} {eT} \frac{\mathcal{L}_{\perp, \|}^{(1)}(\mu,T)} {\mathcal{L}_{\perp, \|}^{(0)}(\mu,T)}
\label{Seeb}
\end{eqnarray}
and the electronic part to the total thermal conductivity accounts to
\begin{equation}
\kappa_{el}{_{\perp, \|}}=\frac{1}{T}(\mathcal{L}_{\perp, \|}^{(2)}(\mu,T)-\frac{(\mathcal{L}_{\perp, \|}^{(1)}(\mu,T))^2}{\mathcal{L}_{\perp, \|}^{(0)}(\mu,T)}) \, .
\label{kel}
\end{equation}
The second term in eq.~\ref{kel} introduces corrections 
due to the Peltier heat flow that can occur when 
bipolar conduction takes place \cite{Tritt:2004p15755}.

The chemical potential $\mu$ at 
temperature $T$ and extrinsic carrier concentration $N$ is determined by an integration 
over the density of states (DOS) $n(E)$
\begin{eqnarray}
N=\int \limits_{-\infty}^{\text{VBM}} \text{d}E \,  n(E) [f(\mu,T)-1]+
\int \limits_{\text{CBM}}^{\infty} \text{d}E \, n(E) f(\mu,T)
\label{Dop},
\end{eqnarray}
where $\text{CBM}$ is the conduction band minimum and $\text{VBM}$ is the 
valence band maximum. 
The k-space integration of eq.~\ref{Tcoeff} for a system with an intrinsic anisotropic texture, e.g. 
in rhombohedral and hexagonal structures, is quite challenging. 
In preceding publications \cite{Yavorsky:2011p15466,Zahn:2011p15523} we stated on the relevance of 
adaptive integration methods needed to reach 
convergence of the energy dependent TDF. 
Especially in regions close to the band edges, which are evident for transport, the anisotropy of 
the TDF requires a high density of the k-mesh. Here, convergence tests 
for the transport properties showed that at least 150 000 k-points in the 
entire BZ had to be included for sufficient high doping rates ($N \geq \unit[1\times 10^{19}]{cm^{-3}}$), while for energies 
near the band edges even more than 56 million k-points were required to reach the 
analytical effective mass values and the corresponding conductivity anisotropies at the band edges. 
%

\section{\label{estruc}Electronic structure}
\begin{figure*}[t]
\centering
\includegraphics[width=0.80\textwidth]{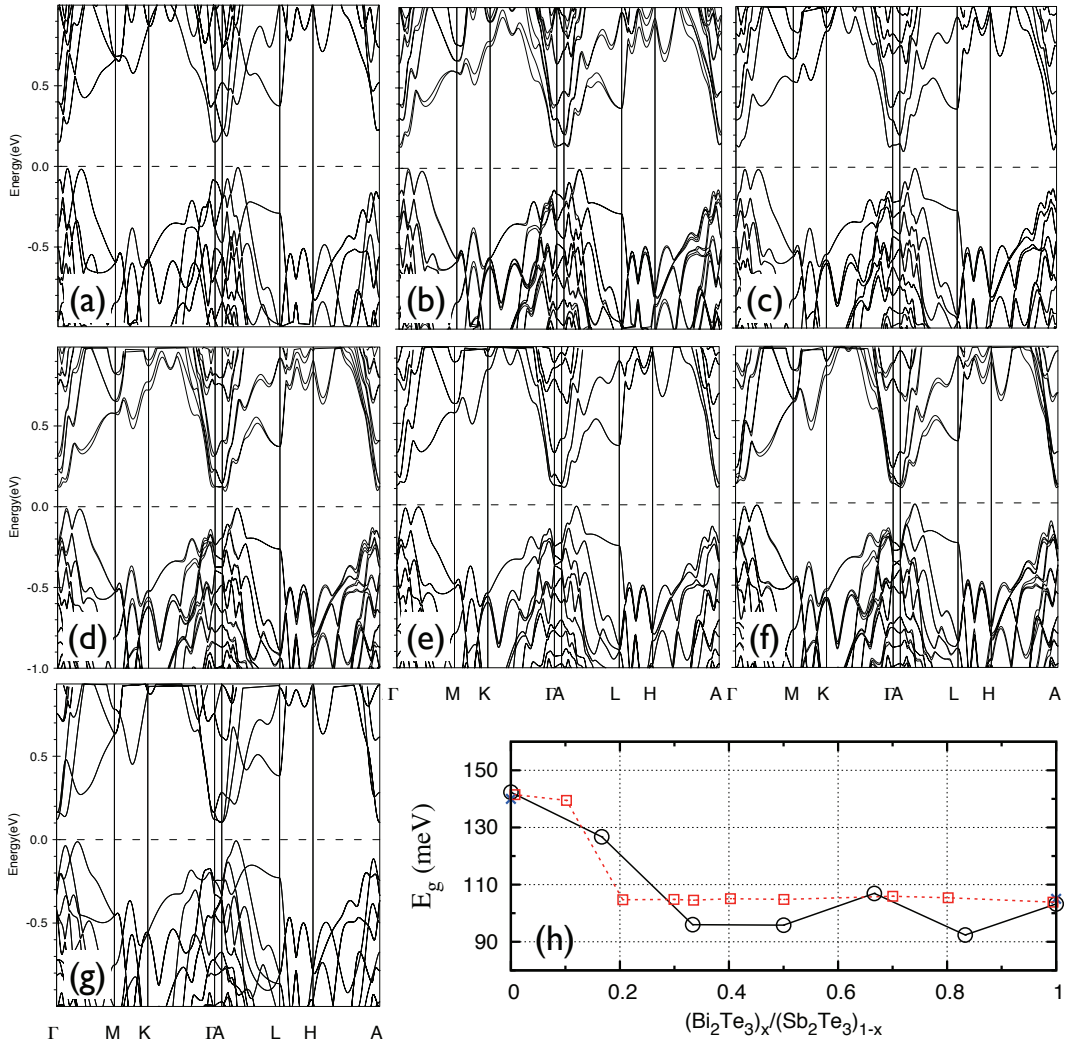}
\caption{\label{fig:2} (color online) Electronic bandstructures 
for $(\text{Bi}_2\text{Te}_3)_{x}/(\text{Sb}_2\text{Te}_3)_{1-x}$ superlattices in the 
hexagonal unit cell with different superlattice periods. (a) $x=0$, (b) $x=\nicefrac{1}{6}$, 
(c) $x=\nicefrac{1}{3}$, (d) $x=\nicefrac{1}{2}$, (e) $x=\nicefrac{2}{3}$, (f) $x=\nicefrac{5}{6}$ 
and (g) $x=1$. In (h) black circles show the calculated fundamental energy gap in dependence on the 
superlattice period. Red squares show experimental findings \cite{Sehr:1962p15527} for 
$(\text{Bi}_2\text{Te}_3)_{x}/(\text{Sb}_2\text{Te}_3)_{1-x}$ alloys, which were linearised to 
allow for comparison with our LDA results.}
\end{figure*}

In \fs{1}(a)-(g) the electronic bandstructures on the hexagonal high symmetry lines 
for all \SBxSLs are shown, starting with (a) $x=0$ 
which is tensile strained bulk \SbTe and ending with  (g) $x=1$, which is bulk \BiTe. 
For the case of (b) $x=\nicefrac{1}{6}$, (d) $x=\nicefrac{1}{2}$ and (f) $x=\nicefrac{5}{6}$ 
a further band splitting can be noticed, which stems from the missing space inversion symmetry in these 
systems and with that the former band degeneracy is lifted. This situation always 
occurs if the Bi(Sb) sites in each quintuple are not uniformly occupied. 

For increasing number of Bi layers in the SLs no drastic change in the band structure 
topology can be stated. Only slight variations were found for the in-plane band directions. 
Of stronger impact could be the change of band dispersion 
which occurs for the lowest lying conduction band in the cross-plane direction $\Gamma$A. Here a continuous 
change of the bands slope is found for increasing amount of Bi layers in the SL. An almost vanishing 
dispersion and very flat bands in cross-plane direction are found for the SL with $x=0.5$, which is three Bi-like 
layers and three Sb-like layers. Further amount of Bi layers in the system leads to an increase in the bands slope, 
while showing different sign compared to bulk \SbTex.

In \f{2}(h) the calculated band gap in dependence on the superlattice period is shown. 
Applying an extended tetrahedron method \cite{Lehmann:1972p14972,Mertig:1987p5922}
and very dense k-mesh's in the BZ, the band gap values were determined within an uncertainty below 1\%. 
While for \SbTe at the experimental lattice parameters, we previously found a direct band 
gap located at the center of the BZ ~\cite{Yavorsky:2011p15466}, an indirect gap of $E_g=\unit[140]{m eV}$ can be stated 
for the in-plane tensile strained \SbTex. For \BiTe as well as for all \SBSLs indirect band gaps 
are obtained, too. 
A known difficulty within standard DFT is the general underestimation of the 
semi-conductors band gaps at zero temperature \cite{MoriSanchez:2008p15799}, as well as the missing 
temperature dependence of $E_g(T)$~\cite{Varshni:1967p14976}. 
For small band gap thermoelectrics, such as \BiTe and \SbTex, this could impinge the TE transport. 
The thermopower might be reduced 
at high temperature and low doping due to bipolar conduction \cite{Singh:2010p14285,Huang:2008p559}. 
This effect would be overestimated if the band gap is underestimated. 
With $E_g(T)$ being considered 
for TE bulk materials\cite{Huang:2008p559,Hinsche:2011p15276} lack of knowledge on the absolute size of the gap, 
as well as its temperature dependence 
permits such gap corrections for the strained bulk materials as well as for the SLs. 
However, the calculated bulk band gap of $E_g=\unit[105]{m eV}$ for unstrained bulk \BiTe is in 
better agreement with the experimental value of $E_g=\unit[130]{m eV}$~\cite{Sehr:1962p15527}, than for unstrained \SbTe where a 
calculated value of $E_g=\unit[90]{m eV}$ faces experimental values between $E_g=\unit[150-230]{m eV}$~\cite{Sehr:1962p15527,vonMiddendorff:1973p15784}. 
As is well known, \BiTe and \SbTe exhibit band inversions at certain areas in the BZ \cite{Zhang:2009p14347}. 
Within LDA the strength of band inversion is most likely underestimated \cite{Yazyev:2012p15832}. At a given band inversion 
the strength of the spin orbit interaction then controls the size of the band gap. Fortuitously, overestimated 
SOC effects and underestimated band inversion tend to cancel each other leading to good results for the 
band gap size and wave function character.

As a lack of data, we can only compare the \SBSL with $x=\nicefrac{1}{3}$ (see \f{2}(c)) 
with previous results of Li \textit{et al.}~\cite{Li:2004p15238}. While they applied a full-potential Linearized 
Augmented Plane Wave (FLAPW) method and treated spin-orbit coupling as a second order perturbation, the results on the bands 
topology are in very good agreement. However, our band gap is substantially 
larger with $E_g=\unit[95]{m eV}$ neglecting any structural relaxations, compared to their value of $E_g=\unit[27]{m eV}$.

\section{\label{unstrained}Thermoelectric transport}
\subsection{Effects of superlattice period composition}
\begin{figure*}[th]
\centering
\includegraphics[width=0.90\textwidth]{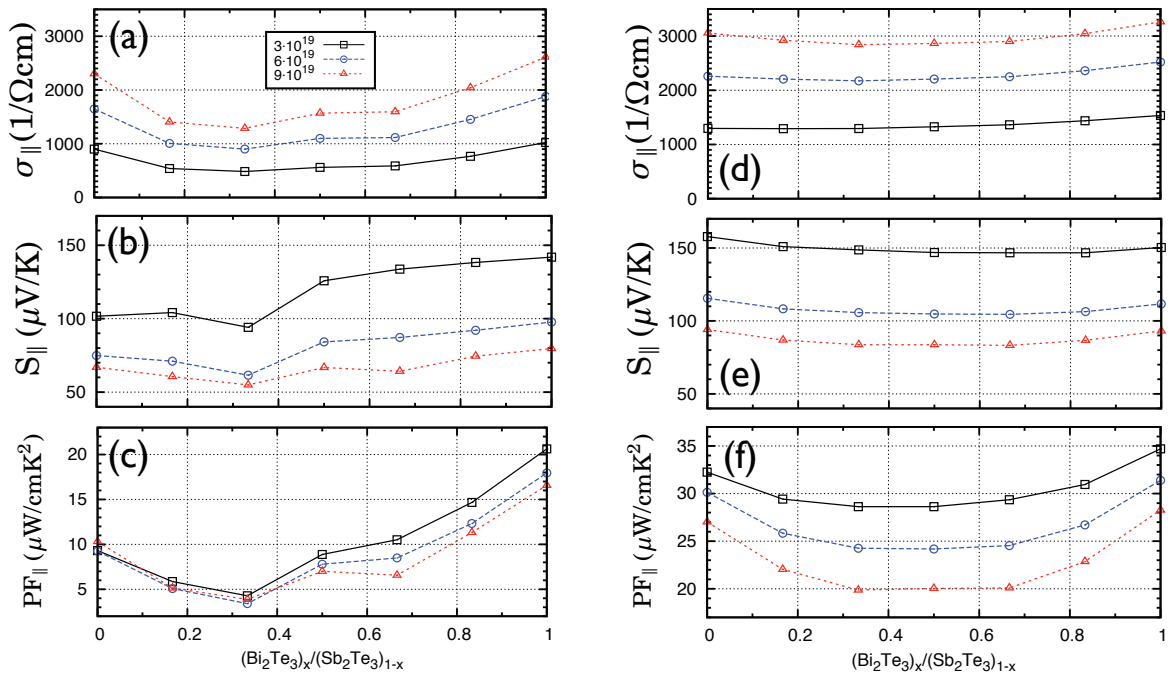}
\caption{\label{fig:3} (color online) Absolute values of in-plane thermoelectric transport properties 
for $(\text{Bi}_2\text{Te}_3)_{x}/(\text{Sb}_2\text{Te}_3)_{1-x}$ superlattices in dependence 
on the superlattice period. Shown are (a),(d) electrical conductivity \cip (b),(e) thermopower \Sip and (c),(f) power factor \PFipx. 
The temperature is fixed to $\unit[300]{K}$ and results for three different charge carrier concentrations (in units of $\unit[]{cm^{-3}}$)
are stated. (a), (b), (c) refer to electron doping, while (d), (e), (f) refer to hole doping.}
\end{figure*}
With knowledge on the electronic structure, we are now able to calculate the desired thermoelectric transport properties 
of the \SBSLsx. As a starting point the in-plane transport properties at room temperature 
for the electron doped (cf. \fs{3}(a)-(c)) and 
hole doped (cf. \fs{3}(d)-(f)) heterostructures are chosen, while afterwards the anisotropy referring to the transport 
in cross-plane direction is discussed in detail. Preliminary studies revealed the optimal charge carrier concentration for the SLs 
to be in the range of $\unit[3-6 \times 10^{19}]{cm^{-3}}$  ~\cite{Hinsche:2011p15707}. For the sake of clarity results are 
presented for three relevant charge carrier concentrations of $3,6$ and $\unit[9 \times 10^{19}]{cm^{-3}}$ 
(cf. solid, dashed and dotted lines in \f{3}, respectively). 

Under electron doping (cf. \f{3}(a)) a decrease of the in-plane electrical conductivity for the superlattices 
compared to the bulk materials is found. This decrease is more pronounced at higher charge carrier concentrations, 
while only slightly being dependent on the SL period. Despite taking into account an isotropic and constant relaxation time 
$\tau = \unit[10]{fs}$, we find very good agreement with experiment \cite{Jeon:1991p13910,Rowe:1995p15325} 
for bulk \BiTe with $\sigma_{_{\|}}=\unit[1030]{(\Omega cm)^{-1}}$ at $N =\unit[3 \times 10^{19}]{cm^{-3}}$. 

The absolute value of the n-type in-plane thermopower is shown in \f{3}(b). At a carrier concentration of 
$N =\unit[3 \times 10^{19}]{cm^{-3}}$ a higher amount of \BiTe in the superlattices leads to a monotonically 
increase in the thermopower from $S{_{\|}}=\unit[103]{\mu V/K}$ (bulk \SbTex) to $S{_{\|}}=\unit[141]{\mu V/K}$
(bulk \BiTex), while showing a dip at a composition of $x=\nicefrac{2}{6}$ with $S{_{\|}}$ below $\unit[100]{\mu V/K}$. 
The latter anomaly is linked to confinement effects and is discussed in detail in Sec.~\ref{QW}. 
This overall behaviour of \Sip is retained for higher charge carrier concentrations at reasonable smaller absolute values. 
Assembling the previous results, the power factor \PFip under relevant electron doping is shown in \f{3}(c). 
Clearly, the reduction of in-plane electrical conductivity \cip, as well as the dip of \Sip at a SL period 
of $x=\nicefrac{2}{6}$ lead to a minimal power factor of about $\unit[4]{\mu W/cm K^{2}}$ at the named 
SL period. We find \PFip for the SLs always to be smaller than expected from an interpolation of 
the bulk absolute values. Furthermore due to compensation effects of \cip and \Sip the 
dependence on the amount of doping is less drastically for \PFip than for it's constituents. The best 
power factor was found for bulk \BiTe to be $\text{PF}_{_{\|}}=\unit[21]{\mu W/cm K^{2}}$, while experimentally 
thin films and single crystals show $\text{PF}_{_{\|}}=\unit[8-27]{\mu W/cm K^{2}}$ and $\text{PF}_{_{\|}}=\unit[45]{\mu W/cm K^{2}}$, 
respectively\cite{Peranio:2011p15471,ScherrerRowe}. 
We note here, that in experiment n-type conduction was only apparent 
for $\text{Bi}_2\text{Te}_3$/$\text{Bi}_2\text{Te}_{2.83}\text{Se}_{0.17}$-SL. 
Nevertheless, to get more insight the physical mechanisms in thermoelectric SL transport, n-type transport 
in \SBSLs should be of enhanced interest, too. 

Highest power factors and FOM were experimentally found for p-type \SBSLsx. The preference for hole conduction is 
dedicated to the large inherent defects introduced by the \SbTe layers. In \fs{3}(d)-(f) the in-plane thermoelectric 
transport properties under hole doping are displayed in the same manner as done before.
Compared to the electron doped case (cf. \f{3}(a)) the hole electrical conductivity \cip is higher at the same charge carrier 
concentration. Furthermore almost no decrease of \cip could be found for the SLs, while this is more visible 
at lower charge carrier concentrations. For the in-plane thermopower the values at different superlattice compositions 
are again only slightly suppressed compared to the bulk systems. For a \SBxSL at $x=\nicefrac{3}{6}$ we state 
$S{_{\|}}=\unit[149]{\mu V/K}$, while $S{_{\|}}=\unit[154]{\mu V/K}$ and $S{_{\|}}=\unit[150]{\mu V/K}$ were 
found for bulk \SbTe and \BiTe at the lowest charge carrier concentration, respectively. This negative bending of the 
thermopower at different superlattice periods is reflected and enhanced for \PFipx. From \f{3}(f) it can be 
seen, that the in-plane power factor \PFip for the various superlattice is decreased compared to the bulk 
materials. However, the largest suppression ($x=\nicefrac{2}{6}$ and $x=\nicefrac{3}{6}$) is found to be 
about 20\% compared to the bulk values, but still offers thermoelectric feasible values 
about $\text{PF}_{_{\|}} =\unit[30]{\mu W/cm K^{2}}$.

To give a reference, in Table \ref{table} the calculated in-plane thermoelectric properties are compared to 
experimental results. In the original work of \V very large values of $\sigma_{_{\|}}$ and $S_{_{\|}}$ result 
in a huge power factor $\text{PF}_{_{\|}}$ about $\unit[72]{\mu W/cm K^{2}}$ at room temperature 
\footnote{We note, that no coherent data of  $N,\sigma_{\|},S_{\|}$ and PF$_{\|}$ for the SL is available. 
$N$ and $\sigma_{\|}$ were concluded from Ref.~\onlinecite{Venkatasubramanian:2001p114}, while 
$S_{\|}$ was estimated from 
Refs.~\onlinecite{Venkatasubramanian:1996p15841,Venkatasubramanian:1997p15831,Venkatasubramanian:1997p15268} 
for comparable SLs, but most probably not from the same sample.}. 
These reported values 
are way larger than found for bulk or thin film \BiTe, \SbTe or their related alloys \cite{Peranio:2011p15471}. 
However, in a more recent study Winkler \textit{et al.}\cite{Winkler:2012p15778} reported values for 
the in-plane electrical conductivity and 
thermopower of a comparable $(\text{Bi}_{0.2}\text{Sb}_{0.8})_2\text{Te}_3/\text{Sb}_2\text{Te}_3$ sputtered SL 
(cf. Table~\ref{tab1}), which are in very good agreement to our theoretical calculations and combine to 
an in-plane power factor $\text{PF}_{_{\|}}$ above $\unit[30]{\mu W/cm K^{2}}$. This is similar to values for bulk
single crystals with comparable compositions. In contrast to the original experiments~\cite{Venkatasubramanian:1997p15268,Venkatasubramanian:2001p114}, which used low-temperature 
metal-organic chemical vapor deposition (MOCVD), Winkler \textit{et al.} applied the concept of ``nano-alloying''~\cite{Konig:2011p48}. 
Here the elemental layers Bi, Sb, and Te are deposited by sputtering and subsequently 
annealed to induce interdiffusion and a solid-state reaction to form the SLs. 
The pronounced periodicity and c-orientation of the SLs have been demonstrated by secondary ion mass spectrometry (SIMS) 
and X-ray diffraction (XRD), respectively.


\begin{table} \footnotesize
\caption{\label{table} Theoretical and experimental in-plane thermoelectric properties of p-type \SBxSLs at room temperature. 
The materials composition amounts in all considered systems to about $x=\nicefrac{1}{6}$. See text for additional details.}
\begin{tabular}{c c c c c c}
\hline\hline N & $\sigma_{\|}$ & $S_{\|}$  & PF$_{\|}$ & Ref.\\
$(\unit[10^{19}]{cm^{-3}})$ & $\unit[]{(\Omega cm)^{-1}}$ & $(\unit[]{\mu V/K})$ & $(\unit[]{\mu W/cm K^{2}})$ \\
\hline
 3.0&$1300$ & $151$ & $30$ & this work\\ 
 3.1&$1818$ & $ \sim 200$ & $ \sim 72$ &  [\onlinecite{Venkatasubramanian:2001p114,Venkatasubramanian:1996p15841,Venkatasubramanian:1997p15831,Venkatasubramanian:1997p15268}] \\ 
 3.2& $761-1160$ & $172-189$ & $27-34$ & [\onlinecite{Winkler:2012p15778}]~\footnote{sputtered $(\text{Bi}_{0.2}\text{Sb}_{0.8})_2\text{Te}_3/\text{Sb}_2\text{Te}_3$ SL} \\ 
 5.8& $3050$ & $115$ & $40$ & [\onlinecite{StordeurRowe}]~\footnote{$(\text{Bi}_{x} \text{Sb}_{1-x})_2\text{Te}_3$ mixed crystal}\\ 
\hline \hline
\label{tab1}
\end{tabular}
\end{table}

\begin{figure*}[]
\centering
\includegraphics[width=0.90\textwidth]{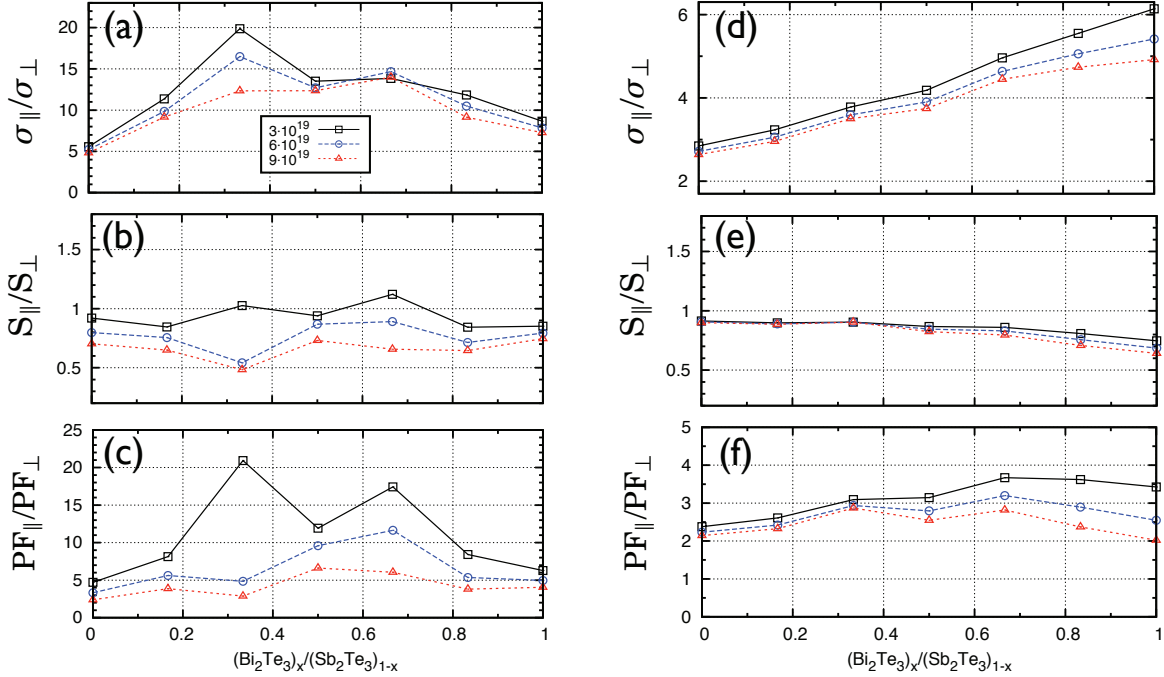}
\caption{\label{fig:4} (color online) Directional anisotropies of thermoelectric transport properties 
for $(\text{Bi}_2\text{Te}_3)_{x}/(\text{Sb}_2\text{Te}_3)_{1-x}$ superlattices in dependence 
on the superlattice period. Shown are (a),(d) electrical conductivity ratio $\nicefrac{\sigma_{\|}}{\sigma_{\perp}}$ 
(b),(e) thermopower ratio $\nicefrac{S_{\|}}{S_{\perp}}$ and 
(c),(f) power factor ratio $\nicefrac{\text{PF}_{\|}}{\text{PF}_{\perp}}$. 
The temperature is fixed to $\unit[300]{K}$ and results for three different charge carrier concentrations (in units of $\unit[]{cm^{-3}}$)
are compared. (a), (b), (c) refer to electron doping, while (d), (e) and (f) refer to hole doping.}
\end{figure*}

While up to now we considered only in-plane transport, in the following the cross-plane transport of the superlattices 
will be discussed. The transport direction is therefore along the SL direction, perpendicular to the hexagonal basal 
plane of the bulk materials. In detail the directional anisotropy of the transport 
properties at room temperature are depicted in \fs{4}(a)-(c) and (d)-(f), 
for electron and hole doping, respectively. To get the absolute values for cross-plane transport, the in-plane values previously 
shown in \f{3} should be divided by the anisotropies presented hereinafter. Anisotropies larger than unity represent 
suppressed thermoelectric transport in cross-plane direction and are therefore less desirable. 
As has been previously proven by experiment \cite{Delves:1961p7491,Jeon:1991p13910,Situmorang:1986p15020,Stordeur:1976p15137,Stordeur:1975p15016} 
and theory\cite{Hinsche:2011p15707,Huang:2008p559,Scheidemantel:2003p14961,Thonhauser:2003p14996}, 
already the bulk thermoelectrics \BiTe and \SbTe show large anisotropies for 
the electrical conductivity, thermopower and the related power factor.

For a sense of purpose, the thermoelectric transport anisotropies under influence of hole doping 
will be considered first. In \f{4}(d) the anisotropy ratio of the electrical conductivity for various 
SL periods is illustrated at a temperature of $\unit[300]{K}$. 
The anisotropy \ratio develops smoothly and monotonously between the bulk limits of 
$\nicefrac{\sigma_{\|}}{\sigma_{\perp}}=2.7$ and about $\nicefrac{\sigma_{\|}}{\sigma_{\perp}}=5-6$ for 
bulk \SbTe and \BiTex, respectively. With increasing amount of \BiTe in the superlattices the 
dependence of \ratio on the charge carrier concentration is more pronounced. This is in 
accordance to previous findings for the bulk materials \cite{Hinsche:2011p15707}.
For the thermopower anisotropy \ratioS this picture holds, too. 
While for \SbTe only a slight anisotropy of about \ratioS=0.9 is found, the asymmetry increases for increasing 
amount of Bi in the SLs, saturating to about \ratioS=0.75 for bulk \BiTex. 
The fact of the cross-plane thermopower being enhanced compared to the in-plane part is well 
known for the two bulk tellurides and compensates somewhat the high electrical conductivity 
anisotropy \ratio to result in a less suppressed cross-plane power factor \cite{Situmorang:1986p15020,Hinsche:2011p15707}. 
The anisotropy for the latter is shown in \f{4}(f). Obviously, \ratioPF is well above unity for all systems indicating 
a less preferred cross-plane electronic transport. Compared to the bulk values of \ratioPF = 2.2 and 
\ratioPF = 2 - 3.5 for bulk \SbTe and \BiTe, respectively, the power factor anisotropy is only slightly 
larger for the SLs with different periods. As an example, for $x=\nicefrac{1}{6}$, which refers to 
a 5{\AA}/25{\AA} \SBSLx, the cross-plane power flow \ratioPF is only suppressed by 13\% with respect to bulk \SbTex, while 
being enhanced by 26\% compared to bulk \BiTe at the optimal charge carrier concentration 
of  $N = \unit[3\times 10^{19}]{cm^{-3}}$. 
For the thermal conductivity in the SLs a suppression compared to bulk 
and the related alloys by about a factor of five is expected \cite{Venkatasubramanian:2000p7305,Venkatasubramanian:2001p114}. 
This would clearly lead to a benefit for the resulting FOM in comparison to bulk, which is discussed more in detail in sec.~\ref{FOM}. 
However, we want to mention, that in the experiments of \Vno a further decrease in the electrical conductivity 
anisotropy was found for thin SLs at various SL periods \cite{Venkatasubramanian:2001p114}. 
It was stated that \ratio under hole doping is about $0.8-1.4$ for the \SBSLs at different SL periods and 
therefore electrical cross-plane transport is strongly 
improved compared to bulk. Our calculations do not show such a trend. 

In \f{4}(a) the electrical conductivity anisotropy \ratio under electron doping is shown. Bulk \BiTe 
and bulk \SbTe show anisotropies around $\nicefrac{\sigma_{\|}}{\sigma_{\perp}}=8$ and 
$\nicefrac{\sigma_{\|}}{\sigma_{\perp}}=5$, respectively, in good agreement to earlier studies\cite{Hinsche:2011p15707}. 
One easily recognizes the anisotropy ratios to be larger than expected from the two bulk limits, 
while obtaining substantially large values of about $\nicefrac{\sigma_{\|}}{\sigma_{\perp}}=20$ at 
$N = \unit[3\times 10^{19}]{cm^{-3}}$ for a SL period 
of $x=\nicefrac{2}{6}$, that is a 10{\AA}/20{\AA} \SBSLx, or one quintuple of \BiTe and two quintuples of \SbTex. 
For the considered case the anisotropy strongly depends on the amount of doping, while decreasing rapidly at 
increased charge carrier concentration, but still reaching 
$\nicefrac{\sigma_{\|}}{\sigma_{\perp}} \ge 10$ at $N = \unit[9\times 10^{19}]{cm^{-3}}$. At the same time the 
thermopower anisotropy shows a clear cross-plane preference at \ratioS$\approx 0.5$. 
Nevertheless, the resulting power factor anisotropy shows disappointing high values of about \ratioPF$\gg 5$ for 
the distinct SLs, while even showing \ratioPFx about 20 for the SL at a composition of $x=\nicefrac{2}{6}$. 
The suppressed cross-plane thermoelectric transport can clearly be linked to the large electrical 
conductivity anisotropies found for the n-type \SBSLsx. In the following we want to emphasize, that 
the latter are related to quantum well effects in the conduction band, which are evoked by a 
conduction band offset between \BiTe and \SbTe in the SLs.

\subsection{\label{QW}Quantum well effects}

In the early 1990's concepts were presented to enhance in-plane thermoelectric properties 
due to the use of quantum-confinement effects in SLs \cite{Hicks:1993p15780,*Hicks:1993p14911,Dresselhaus:1999p15757}. 
While huge enhancements on the in-plane figure of merit were predicted, the authors suppressed 
electron tunnelling and thermal currents between the layers by introducing infinite potential barriers and zero 
barrier widths. Later on it was shown, that for realistic barrier heights and widths the 
enhancement is rather moderate, predicting ZT values that at theirs best are a few percent larger than 
corresponding bulk materials\cite{Sofo:1994p15420,Broido:1995p15758}.

It is known for the two tellurides, that due to SOC induced band inversions 
near the $\Gamma$ point the CBM is derived from states localized at the Te atoms whereas 
the VBM is formed by Bi or Sb orbitals \cite{Zhang:2009p14347,Eremeev:2012p15782}. 
Nevertheless, contributions from other areas in the BZ lead to the fact, that the valence band conduction contribution 
is mainly determined by Te states at appropriate charge concentrations. 
Due to this, a substitution of Bi and Sb in \BiTe or \SbTe affects the character of the valence band 
states only marginally and almost bulk-like electronic structure and transport 
properties can be expected in the SLs.

\V argued that in the SLs due to weak-confinement and near-zero band-offset, 
there is minimal anisotropy between in-plane and cross-plane electrical conductivities. 
While we can confirm that the valence band offset is almost vanishing in all SLs, the argumentation of 
\Vline would conclude, that in bulk \BiTe and \SbTe \ratio$=1$, as the band offsets in 
bulk materials are zero by definition\cite{Sootsman:2009p15788}. The latter conclusion is obviously not the case.

\BiTe and \SbTe show a theoretical band gap difference of about $\unit[35]{meV}$, 
thus a band offset in the SL is expected. Our calculations reveal 
that this difference is mostly located in the conduction bands. 
The offset in the conduction band edges of \BiTe and \SbTe sets 
up potential barriers in the superlattice, which leads to confinement of the electrons in the well regions.
Therefore, the site resolved probability amplitude for the two bulk tellurides, as well as for the 
superlattice which showed the highest conductivity anisotropy, i.e. 10{\AA}/20{\AA} \SBSL ($x=\nicefrac{2}{6}$), 
is shown in \F{5}(a)-(c). 

\begin{figure}[]
\centering
\includegraphics[width=0.48\textwidth]{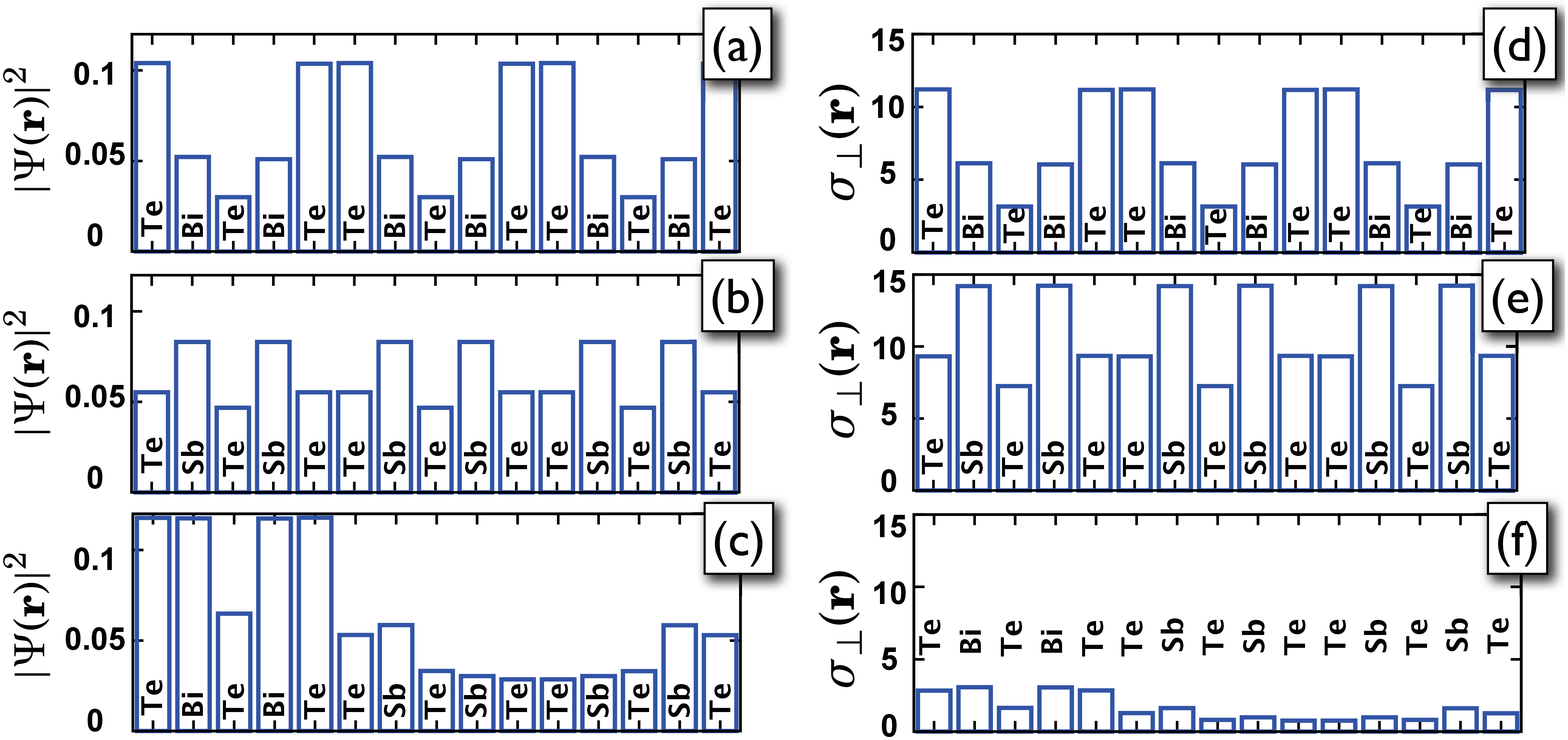}
\caption{\label{fig:5} (color online) Site resolved probability amplitude for (a) bulk \BiTe, (b) bulk 
\SbTe and (c) a $(\text{Bi}_2\text{Te}_3)_{x}/(\text{Sb}_2\text{Te}_3)_{1-x}$ superlattice with $x=\nicefrac{2}{6}$.
In the same manner the site resolved cross-plane electrical conductivity \cpp (in units of $\unit[]{(\Omega cm)^{-1}}$) is shown for (d) bulk \BiTe, (e) bulk 
\SbTe and (f) a $(\text{Bi}_2\text{Te}_3)_{x}/(\text{Sb}_2\text{Te}_3)_{1-x}$ superlattice with $x=\nicefrac{2}{6}$. 
The temperature is fixed to $\unit[300]{K}$ and the charge carrier concentration is set to $N = \unit[3\times 10^{19}]{cm^{-3}}$.}
\end{figure}

For \BiTe the conduction band edge is dominated by contributions of the Te$_1$ followed by the Bi sites. A 
localization of states nearby the \textsc{van der Waals} gap is already visible in the bulk system. 
The fact, that the band inversion does not heavily affect 
the orbital character at the CBM is caused by the indirect band gap character of \BiTex. Areas of the BZ 
where no band inversion occurs dominate the CBM.

For \SbTe this is quite different. Even though \SbTe changes from an direct to an indirect semiconductor under applied 
in-plane tensile strain \cite{Yavorsky:2011p15466}, the CBM remains nearby the $\Gamma$ point in the BZ. 
Therefore most of the contributions to conductivity arise from an area around the $\Gamma$ point, where the band inversion 
favours the Sb character. As can be seen from \F{5}(b) the contributions to the DOS are equally distributed 
over all positions in the unit cell, while slightly being enlarged on the Sb sites.
For the \SBxSL at $x=\nicefrac{2}{6}$ (cf. \F{5}(c)) we found quantum well states, which localize 
about half of the density in the \BiTe quintuple, while the density in the two \SbTe quintuples 
is strongly depleted. 
We obtain similar results, if two quintuples ($x=\nicefrac{4}{6}$) are occupied by \BiTex. 

As shown in \F{5}(f) this quantum confinement is reflected in the contribution to the cross-plane 
electrical conductivity. Here, the local cross-plane conductivity $\sigma_{\perp}(\bold{r})$ is calculated 
as introduced in Ref.~\onlinecite{Zahn:1998p15803}, by weighting the contributions to $\mathcal{L}_{\perp, \|}^{(0)}(\mu, T)$ 
with the normalized probability amplitude $\left|\Psi(\bm{r})\right|^{2}$ of the electronic states at chemical potential $\mu$. 
Summing up $\sigma_{\perp}(\bold{r})$ over all sites gives the total electrical conductivity $\sigma_{\perp}$ (cf. \f{3}(a) and \f{4}(a)). 
Weighting the DOS $n(\mu)$ with the normalized probability amplitude $\left|\Psi(\bm{r})\right|^{2}$ refers 
to the local DOS $n(\mu,\bm{r})$ (LDOS). 

Nevertheless sites in the \BiTe quintuple with more accumulated density 
carry a larger contribution to the conductivity \cppx, the total contribution compared to the bulk tellurides 
is strongly suppressed (comp. \F{5}(d),(e)). This can be affirmed by a picture that electrons travelling 
in the cross-plane direction are exposed to a tunneling-like behaviour for about the distance of the \SbTe 
quintuples. This clearly leads to a diminished cross-plane group velocity of the electronic states. 
Comparing \F{5}(d) and (e) we see furthermore  that even the localization inside the quintuple in 
bulk \BiTe can lead to reduced cross-plane electronic transport, reflected 
in larger total anisotropies about $\nicefrac{\sigma_{\|}}{\sigma_{\perp}}=8$ for bulk \BiTe compared 
to $\nicefrac{\sigma_{\|}}{\sigma_{\perp}}=5$ for bulk \SbTex.

To extend to the results obtained at room temperature in \F{6} the temperature dependence of 
the in-plane and cross-plane thermopower and power factor are presented at a electron/hole 
charge carrier concentration of $N = \unit[3\times 10^{19}]{cm^{-3}}$. 
The p-type thermopower shows only moderate dependencies on the SL period at all 
temperatures with the anisotropy slightly favouring the cross-plane part \Sppx. Under electron 
doping the dependence of the thermopower on the SL period is more pronounced, which can 
to some amount be assigned to the quantum well effects which occur in the conduction band. At higher 
temperatures the thermopowers anisotropy is distinct larger. The latter was shown 
before for the bulk materials \cite{Hinsche:2011p15707}. Due to the fact of the thermopower, 
as well as the electrical conductivity being clearly smaller under electron doping than 
hole doping (cf. \F{3}) we find the largest values of PF for the p-type SLs. Here the largest 
was found to be $\text{PF}_{_{\|}}=\unit[42]{\mu W/cm K^{2}}$ at $\unit[500]{K}$ 
for \SBxSL with $x=\nicefrac{1}{6}$. For bulk \BiTe and \SbTe we state maximum 
$\text{PF}_{_{\|}}=\unit[44]{\mu W/cm K^{2}}$ and $\unit[49]{\mu W/cm K^{2}}$ 
at $\unit[400]{K}$ and $\unit[550]{K}$, respectively. Due to 
previously discussed conductivity anisotropy \ratiox, the cross-plane PF is 
strongly suppressed and the maxima are shifted to higher temperatures. 

\begin{figure}[]
\centering
\includegraphics[width=0.48\textwidth]{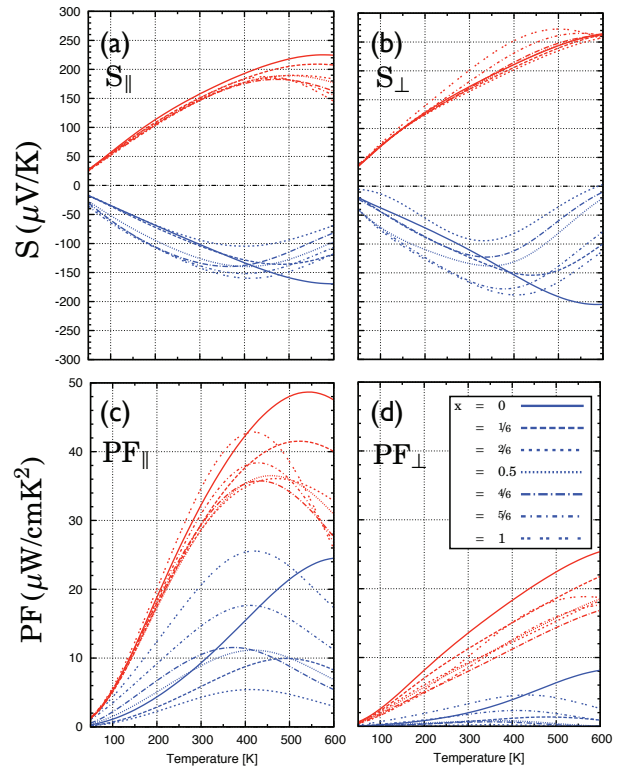}
\caption{\label{fig:6} (color online) Temperature dependence of thermoelectric transport properties 
for $(\text{Bi}_2\text{Te}_3)_{x}/(\text{Sb}_2\text{Te}_3)_{1-x}$ superlattices. 
Shown are (a) in-plane thermopower \Sip and (b) cross-plane thermopower \Sppx, as well as 
the corresponding power factors (c)  \PFip in the in-plane and (d) \PFpp in the cross-plane direction. 
The charge carrier concentration is fixed to $N = \unit[3\times 10^{19}]{cm^{-3}}$ and 
different line types correspond to different superlattice periods. 
Blue lines refer to electron doping, while red lines refer to hole doping.}
\end{figure}

%
%
\subsection{\label{FOM}Towards figure of merit}

With the electronic transport properties discussed in the previous sections, we are now 
going to focus on the electronic and lattice part contribution 
to the thermal conductivity $\kappa_{el} + \kappa_{ph}$ to give some estimations on the FOM. 
As has been stated before, the main benefit from a superlattice structure for the FOM 
is expected from a reduction of the cross-plane thermal conductivity at retained electronic transport properties. 
Today, the reduction of the cross-plane lattice thermal conductivity in 
thermoelectric superlattices has been widely and successfully proven 
\cite{Lee:1997p1545, BorcaTasciuc:2000p15132,Huxtable:2002p2508,Chakraborty:2003p15514}.

In the past thermal conductivity reduction in crystalline or polycrystalline 
bulk thermoelectric materials was traditionally achieved by alloying. 
However, one reaches the so-called “alloy limit” of thermal conductivity, 
which has been difficult to surpass by nanostructuring \cite{Vineis:2010p14995}.

Nevertheless, for \SBSLs cross-plane lattice thermal conductivites of $\kappa_{ph}=\unit[0.22]{W/m K}$ 
were reported for certain SL periods, which is a factor of two below 
the alloy limit \cite{Venkatasubramanian:2000p7305}. It is obvious, that at 
thermoelectric relevant charge carrier concentrations and temperature ranges, the 
electronic contribution $\kappa_{el}$ can be in the same order of magnitude. 

\begin{figure}[]
\centering
\includegraphics[width=0.48\textwidth]{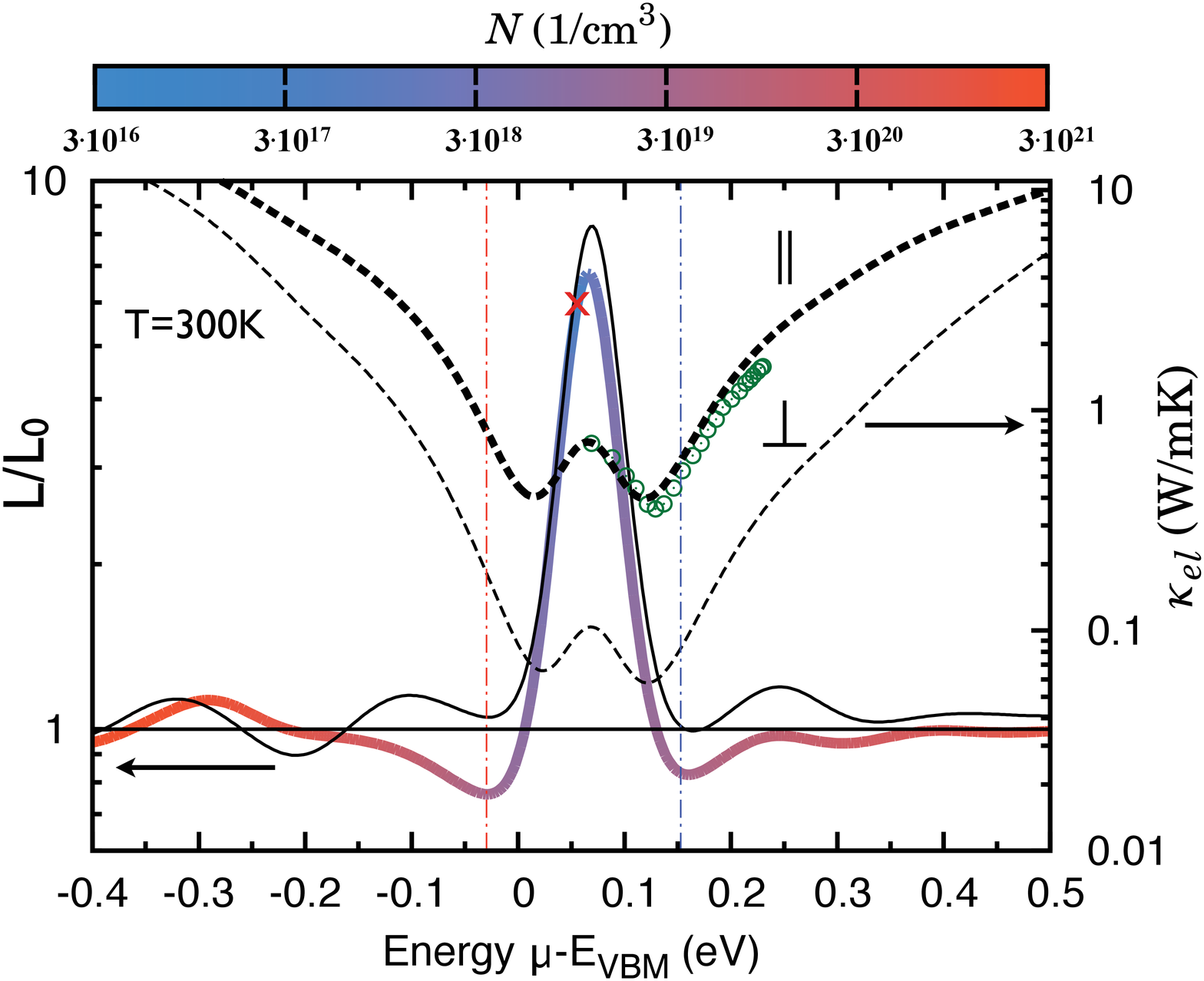}
\caption{\label{fig:7} (color online) Lorenz function L (solid lines, ref. to left scale) and 
electronic contribution $\kappa_{el}$ to the total thermal conductivity (dashed lines, ref. to the right scale) 
in dependence on position of the chemical potential $\mu$ for 
bulk $\text{Bi}_2\text{Te}_3$ for the in-plane (thick lines) 
and cross-plane (thin lines) transport direction. The Lorenz function is related to the 
metallic limit $L_0=\unit[2.44\times 10^{-8}]{W \Omega/K^{2}}$. 
Plotted on to the Lorenz function 
in the in-plane direction is a color code referring to the charge carrier concentration 
given by Eq.~\ref{Dop}. The red cross emphasizes the change from n to p doping,
respectively. 
The temperature was fixed to $\unit[300]{K}$. Thin vertical dash-dotted lines 
emphasize the position of the chemical potential for a 
charge carrier concentration of $N = \unit[3\times 10^{19}]{cm^{-3}}$ under 
n and p doping (blue and red line, respectively). The CBM is located at $\unit[0.105]{eV}$. 
Green open circles show experimental values from Ref.~\cite{Goldsmid:1965p15735} 
for $\kappa_{el,\|}$ for an n-type \BiTe single crystal.}
\end{figure}

Therefore \F{7} shows the room temperature doping dependent electronic part 
of the thermal conductivity, in the in-plane (thick dashed lines, right scale) and 
cross-plane direction (thin dashed lines, right scale), for bulk \BiTex, 
to give insight in the principle dependencies. Furthermore, the Lorenz function 
defined via Eqs.~\ref{Seeb} and \ref{kel} as 
$L{_{\perp, \|}}=\kappa_{el}{_{\perp, \|}} \cdot (\sigma{_{\perp, \|}} \cdot T)^{-1}$
is shown for the in-plane (thick solid line, left scale and color code) 
and cross-plane part (thin solid line, left scale), respectively. 
As can be seen, $\kappa_{el} $ minimizes for energies near the band edges. Here, 
at $N \approx \unit[3\times 10^{18}]{cm^{-3}}$, the thermopower S maximizes at 
appropriate values for the electrical conductivity $\sigma$, hence the second term 
in Eq.~\ref{kel} increases leading to small values for $\kappa_{el}$. At small intrinsic 
charge carrier concentrations, the chemical potential shifts into the gap and 
the total thermopower is strongly reduced due to 
bipolar diffusion. This leads to an enhanced contribution to the electrical 
thermal conductivity at intrinsic charge carrier concentrations and is known 
as the bipolar thermodiffusion effect \cite{Frohlich:1954p15802,Goldsmid:1965p15735,Uher:1974p15736}. 
At charge carrier concentrations of $N = \unit[3\times 10^{19}]{cm^{-3}}$ 
we find $\kappa_{el}{_{\|}} $ to be about $\unit[0.6-0.8]{W/m K}$ for n/p-type 
bulk $\text{Bi}_2\text{Te}_3$ in very good agreement 
with experimental (cf. green, open circles in \f{7}) 
and theoretical results \cite{Peranio:2006p15247,Goldsmid:1956p15499,Huang:2008p559}. 
The cross-plane component of $\kappa_{el}$ 
is substantially smaller, especially for n-type conduction, reflecting here the electrical 
conductivity anisotropy discussed earlier. 
The bipolar thermodiffusion effect is furthermore responsible for the suppression of the Lorenz function to values 
below the metallic limit $L_0$ ($L_0=\unit[2.44\times 10^{-8}]{W \Omega/K^{2}}$) 
for values of the chemical potential near the band edges (cf. \F{7} solid lines, right scale). 
At optimal charge carrier concentrations of $N = \unit[3\times 10^{19}]{cm^{-3}}$ 
$L_{\|} \approx 0.7 L_0$ under hole doping (red dashed dotted lines) and 
$L_{\|} \approx 0.8 L_0$ under electron doping (blue dashed dotted lines) can be found. 
For the cross-plane Lorenz function $L_{\perp} \approx L_0$ is stated at the same 
amount of n/p-type doping. Reaching the intrinsic doping regime the Lorenz function 
reaches substantially large values of $L_{\|} \approx 6.5 L_0$ and $L_{\perp} \approx 8 L_0$. 
Such a behaviour has been described in literature \cite{Chaput:2005p1405,Huang:2008p559} 
and can have consequences for the determination of the thermal conductivity. 
The Lorenz factor is generally used to separate $\kappa_{el}$ and $\kappa_{ph}$. 
At thermoelectric advisable charge carrier concentrations applying the metallic value 
$L_0$ to determine the lattice thermal conductivity could lead to an overestimation 
of the electronic thermal conductivity, and consequently to underestimation 
of the lattice contribution. The Lorenz function of thermoelectric heterostructures can 
show further anomalies which are discussed in detail in a forthcoming publication \cite{HinscheICT2012}. 

\begin{figure*}[t]
\centering
\includegraphics[width=0.90\textwidth]{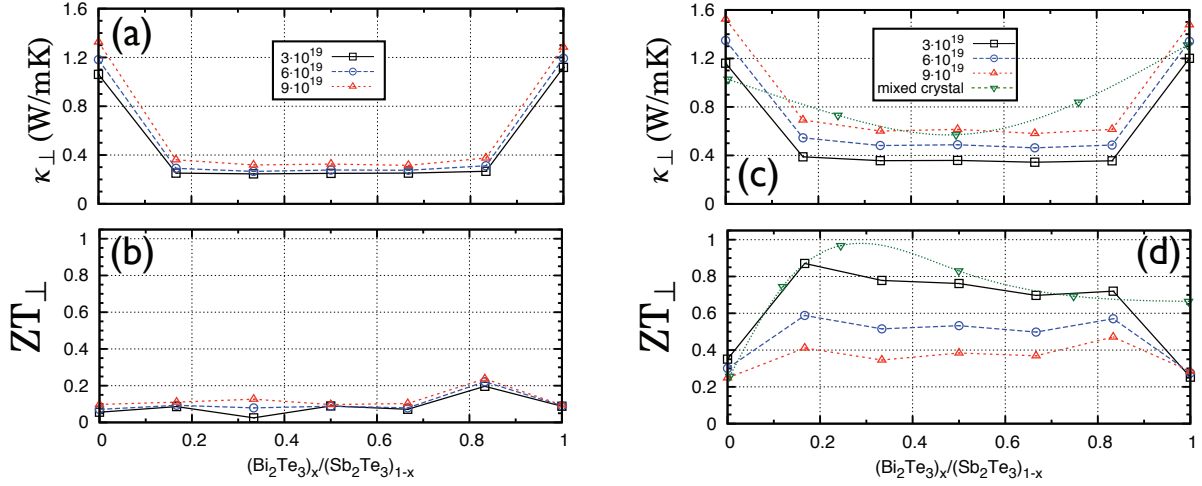}
\caption{\label{fig:8} (color online) Absolute values of cross-plane thermoelectric transport properties 
for $(\text{Bi}_2\text{Te}_3)_{x}/(\text{Sb}_2\text{Te}_3)_{1-x}$ superlattices in dependence 
on the superlattice period. Shown are (a),(c) the total thermal conductivity $\kappa_{el}+\kappa_{ph}$ 
and (b),(d) the cross-plane figure of merit ZT$_\perp$. 
The temperature is fixed to $\unit[300]{K}$ and results for three different charge carrier concentrations 
are compared. (a), (b) refer to electron doping, while (c), (d) refer to hole doping. The electronic part $\kappa_{el}$ was 
calculated, while the lattice part $\kappa_{ph}$ was taken from literature 
\cite{Venkatasubramanian:2000p7305,Venkatasubramanian:2001p114}. As a reference point experimental 
results for the mixed single crystal series $(\text{Bi}_{x} \text{Sb}_{1-x})_2\text{Te}_3$ at comparable 
material composition are shown as green downward triangles \cite{StordeurRowe}.}
\end{figure*}

Experimental findings for the lattice part $\kappa_{ph}$ of the thermal conductivity are added 
to the calculated electronic contribution $\kappa_{el}$ to present some estimations 
on the cross-plane FOM ZT$_{\perp}$. In particular 
$\kappa_{ph,\perp}=\unit[1.05]{W/m K}$, $\kappa_{ph,\perp}=\unit[0.96]{W/m K}$ and 
$\kappa_{ph,\perp}=\unit[0.22]{W/m K}$ at room temperature were used for bulk \BiTex, \SbTe 
and the \SBSLs \cite{Venkatasubramanian:2000p7305}, respectively \footnote{We want to mention that the 
lattice thermal conductivity was found to be a function of the SL period 
\cite{Venkatasubramanian:2000p7305}. As a lack of further data, we assumed the smallest $\kappa_{ph,\perp}$ 
for all of our SL periods.}. 

Recently Winkler \textit{et al.}\cite{Winkler:2012p15778} measured 
for a p-type $(\text{Bi}_{0.2}\text{Sb}_{0.8})_2\text{Te}_3/\text{Sb}_2\text{Te}_3$ SL 
the total cross-plane thermal conductivity $\kappa_{\perp}$ 
to be about $\unit[0.45-0.65]{W/m K}$ for different annealing temperatures. 
The values were obtained within a time-domain thermal reflectance (TDTR) measurement and are 
in very good agreement to our calculations, which are displayed in \F{8}(c). Compared to the 
original experiments by \Vx, the values of the total cross-plane thermal conductivity 
are smaller. This stems to a large extent from the fact that a strong electrical conductivity anisotropy 
\ratio is apparent and not vanishing to $\nicefrac{\sigma_{\|}}{\sigma_{\perp}} \sim 1$ as proposed in \Vlinex. 
Hence not only \cpp but also $\kappa_{el,\perp}$ is 
noticeably suppressed. Furthermore in Refs.~\onlinecite{Venkatasubramanian:2001p114,Venkatasubramanian:2000p7305} 
it was suggested, that mirror-like SL interfaces lead to potential reflection effects 
and reduce $\kappa_{ph,\perp}$ very efficiently. Beside that, Touzelbaev \textit{et al.}~\cite{Touzelbaev:2012p15270} 
showed that an existing interface roughness 
will additionally decrease $\kappa_{ph,\perp}$. Beside the nano-crystallinity of the samples, 
such an additional interface roughness is most likely provided by interdiffusion effects at the interfaces introduced by the growth 
of \SBSLs within the concept of ``nano-alloying''~\cite{Konig:2011p48,Winkler:2012p15778}.

In \F{8} the room-temperature cross-plane properties of the total thermal conductivity and the related cross-plane part 
of the FOM are shown, for electron (a),(b) and hole doping (c),(d), respectively. As could be 
expected the strong quantum well effects in the conduction band lead to quite small values for 
ZT$_{\perp} \le 0.2$ under electron doping. Here a reduction of the SL total thermal conductivity of about a factor 5-6 
compared to bulk is impeded by an accompanied reduction of PF$_{\perp}$ by a factor of 10-20 (cf. \F{4}(c)). Thus 
no benefit for the FOM in the cross-plane direction can be revealed under electron doping. 
At hole doping the situation is more advantageous. Beside the total thermal conductivity (cf. \F{8}(c) and \F{7}) being somewhat 
larger than under electron doping, the electronic transport properties remain bulk-like for all SL periods (cf. \F{4}(f)) and 
hence an enhancement for ZT$_{\perp}$ by a factor of 2-3 can be achieved under hole doping of 
$N = \unit[3\times 10^{19}]{cm^{-3}}$. We state values ZT$_{\perp} \approx 0.7-0.9$ for different SL periods, while 
an higher amount of \SbTe in the SL leads to larger values for ZT, despite no distinct influence of the SL period 
on the FOM could be found. As a supplement experimental data \cite{StordeurRowe} 
for $\kappa_{\perp}$ and ZT$_{\perp}$ of 
the mixed single crystal series $(\text{Bi}_{x} \text{Sb}_{1-x})_2\text{Te}_3$ 
is shown as green downward triangles in \F{8}(c) and (d). The hole doping varied steadily between $N = \unit[1\times 10^{19}]{cm^{-3}}$ 
for \BiTe to $N = \unit[9\times 10^{19}]{cm^{-3}}$ for \SbTex. One easily abstracts that ZT$_{\perp}$ values 
of the optimal doped SLs and the mixed single crystal 
series show clear similarities considering the dependence on the materials composition, as well as the the absolute values of ZT$_{\perp}$.

\begin{figure}[]
\centering
\includegraphics[width=0.48\textwidth]{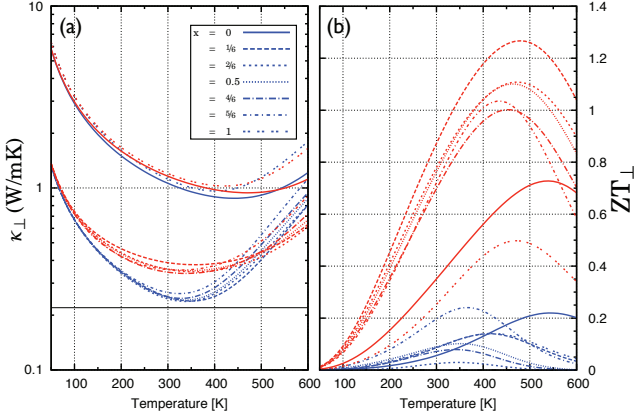}
\caption{\label{fig:9} (color online) Temperature dependence of the thermal 
conductivity and figure of merit for the 
$(\text{Bi}_2\text{Te}_3)_{x}/(\text{Sb}_2\text{Te}_3)_{1-x}$ SLs. 
Shown are (a) total cross-plane thermal conductivity $\kappa_{\perp}=\kappa_{el,\perp}+\kappa_{ph,\perp}$ and (b) cross-plane figure of merit. 
The charge carrier concentration is fixed to $N = \unit[3\times 10^{19}]{cm^{-3}}$ and 
different line types correspond to different SL periods. 
Blue lines refer to electron doping, while red lines refer to hole doping. The electronic part $\kappa_{el}$ was 
calculated, while the lattice part $\kappa_{ph}$ was taken from literature 
\cite{Venkatasubramanian:2000p7305,Venkatasubramanian:2001p114}.}
\end{figure}

To extend our findings at room-temperature, in \F{9}(a) and (b) 
temperature dependent results for the total cross-plane thermal conductivity and cross-plane 
FOM are shown for the \SBxSLs at electron/hole concentration of 
$N = \unit[3\times 10^{19}]{cm^{-3}}$. Within, a conventional $\nicefrac{1}{T}$ 
dependence for the lattice thermal 
conductivity was assumed \cite{Peranio:2008p15241}, while the calculated electronic part $\kappa_{el}$ is temperature and 
doping dependent, per se. 
We note that for thermoelectric SLs no clear tendency on the temperature dependence 
of $\kappa_{ph}$ can be revealed. However, conventional $\nicefrac{1}{T}$ dependence, 
as well as temperature independent $\kappa_{ph}$ were found experimentally \cite{BorcaTasciuc:2000p15132,Hase:2011p15728,Lee:1997p1545}. 
Models show, that $\kappa_{ph}$ should diminish at low periods\cite{Pattamatta:2009p15730}, 
while experiments reveal a saturation towards the alloy limit for SL periods below 50\AA \cite{Venkatasubramanian:2000p7305}.

As can be seen from \F{9}(a) $\kappa_{\perp}$ takes a minimum at  about $\unit[300-400]{K}$ for the \SBSLsx. 
This behaviour is dictated by the electronic contribution to $\kappa_{\perp}$ and 
supported by the $\nicefrac{1}{T}$ dependence of the lattice part of $\kappa$. 
At low temperature 
the chemical potential is located in the bands an thus a moderate contribution to $\kappa_{el,\perp}$ 
is obtained. With increasing temperature $\kappa_{el,\perp}$ slightly decreases as the chemical potential 
shifts towards the band edges, then reaching minimal $\kappa_{el,\perp}$ for chemical potential 
positions at the band edges (cf. \F{7}). At elevated temperatures the bipolar contribution leads to 
an enhanced electronic contribution to $\kappa_{\perp}$, which then clearly dominates the 
$\nicefrac{1}{T}$ dependence of the lattice part of $\kappa$ leading to large values of the 
total thermal conductivity. The influence of the electronic contribution is more pronounced 
in the SLs compared to bulk, as here $\kappa_{el,\perp} \geq \kappa_{ph,\perp}$. 
Combining these results with the temperature dependent power factor \PFpp discussed in \F{5}(d) 
we find the temperature dependence on the cross-plane FOM as presented in \F{9}(b). 
Concentrating on the more promising p-type SLs we state maximized values for ZT$_{\perp}$ 
clearly above unity for temperatures of $\unit[400-500]{K}$. The largest cross-plane 
FOM is found to be ZT$_{\perp}=1.27$ at about $\unit[470]{K}$ for a \SBxSL at $x=\nicefrac{1}{6}$ 
and a hole concentration of $N = \unit[3\times 10^{19}]{cm^{-3}}$. 
We want to mention that in the experiments of \Vline the maximal ZT$_{\perp}$ under hole 
doping was stated at a SL period of $x=\nicefrac{1}{6}$, too. 
The best value for an n-type SL is ZT$_{\perp}=0.25$ at about $\unit[360]{K}$ at a 
SL period of $x=\nicefrac{5}{6}$.

\section{Conclusion} 

The anisotropic thermoelectric transport 
properties of \SB superlattices at different superlattice periods is presented to get insight into the 
physical mechanisms which are responsible for the path-breaking experimental results with 
ZT$_{\perp}=2.4$ at room temperature obtained by \Vlinex. 
Several aspects added up to obtain those very high ZT values in experiment. 
\textit{(i)} In-plane values of the electrical conductivity and thermopower were found to be larger 
than in the bulk systems at comparable charge carrier concentrations. 
\textit{(ii)} An elimination of the electrical conductivity anisotropy \ratiox, which is apparent in both 
bulk systems, was found in superlattices at certain periods. 
\textit{(iii)} The lattice part of the thermal conductivity was reduced below the alloy limit due to 
phonon-blocking at the superlattice interfaces \cite{Venkatasubramanian:2000p7305,Touzelbaev:2012p15270}. 

Even though taking into account the most optimistic value $\kappa_{ph,\perp}=\unit[0.22]{W/m K}$ 
for the lattice part of the thermal conductivity, we found the cross-plane figure of merit for the best 
p-type superlattice to be ZT$_{\perp}=0.9$ at room temperature and slightly 
enhanced to ZT$_{\perp}=1.27$ at elevated temperature. However, this is more than a factor of two worse than 
experimentally revealed and is caused by the fact, that within the presented \textit{ab initio} 
calculations the findings \textit{(i)} and \textit{(ii)} could not be confirmed. We want to add, 
that in first consistent experiments by \Vno~\cite{Venkatasubramanian:1996p15841} a room-temperature 
ZT$_{\perp}=1.2$ was proposed for a non-symmetrical \SB superlattice with a period of 30\AA.

For the in-plane transport properties of S, $\sigma$ and PF we can state values comparable to bulk 
for the p-type and n-type superlattices, which is in agreement to recent experiments \cite{Winkler:2012p15778,Konig:2011p48}. 
Furthermore for the p-type superlattices a conservation of the bulk transport anisotropies is found, but in no case a reduction, 
while under electron doping strong 
quantum well effects due to conduction band offsets lead to large transport anisotropies \ratio$\geq 10$ and 
suppress the cross-plane thermoelectric transport notably.

Concluding, the experimentally found remarkable thermoelectric transport properties in \SB superlattices 
could not be revealed by detailed band structure effects. An ongoing issue will be to clarify, 
whether scattering effects caused by defects and lattice imperfections could give a considerable 
leap forward to understand the enhanced thermoelectric efficiency in \SB superlattices.

\begin{acknowledgments}
  This work was supported by the Deutsche
  Forschungsgemeinschaft, SPP 1386 `Nanostrukturierte Thermoelektrika: 
  Theorie, Modellsysteme und kontrollierte Synthese'. N. F. Hinsche is
  member of the International Max Planck Research School for Science
  and Technology of Nanostructures. 
\end{acknowledgments}

\bibliography{draft_2.bbl}

\end{document}